# Maxwell Equations without a Polarization Field

## using a paradigm from biophysics

**Robert S. Eisenberg***

Department of Applied Mathematics
Illinois Institute of Technology
Chicago IL 60616 USA
Reisenberg@iit.edu

Department of Physiology and Biophysics
Rush University Medical Center
Chicago IL 60612 USA
Beisenbe@rush.edu

Correspondence: bob.eisenberg@gmail.com
Tel.: +01-708-932-2597

File name:

Maxwell Equations without a Polarization Field September 16 2020.docx

# Abstract

Electrodynamics is usually written using polarization fields to describe changes in distribution of charges as electric fields change. This approach does not specify polarization fields uniquely from electrical measurements.

Many polarization fields will produce the same electrodynamic forces and flows because only divergence of polarization enters Maxwell's equation, relating charge and electric field. The curl of any function can be added to a polarization field without changing the electric field at all. The divergence of the curl is always zero. Models must describe the charge distribution and how it varies to be unique.

I propose a different paradigm to describe field dependent charge, i.e., the phenomenon of polarization. This operational definition of polarization has worked well in biophysics for fifty years, where a field dependent, time dependent polarization provides gating current that makes neurons respond sensitively to voltage. Theoretical estimates of polarization computed with this definition fit experimental data.

I propose that operational definition be used to define polarization charge in general. Charge movement needs to be computed from a combination of electrodynamics and mechanics because 'everything interacts with everything else'. The classical polarization field need not enter into that treatment at all.

When nothing is known about polarization, it is necessary to use an approximate representation with a dielectric constant that is a single real positive number. This approximation allows important results in some cases, e.g., design of integrated circuits in silicon semiconductors, but can be seriously misleading in other cases, e.g., ionic solutions.

## 1. Introduction

Polarization has a central role in electrodynamics. Faraday and Maxwell thought all charge depends on the electric field. All charge would then be polarization.

Maxwell used the **D** and **P** fields as fundamental dependent variables. Charge only appeared as polarization, usually over-approximated by a dielectric constant $\varepsilon_r$ that is a single real positive number. Charge independent of the electric field was not included, because the electron had not been discovered: physicists at Cambridge University (UK) did not think that charge could be independent of the electric field. The electron was discovered some decades later, in Cambridge, ironically enough [1, 2][1]. It then became apparent to all that the permanent charge of an electron is a fundamental source of the electric field. The electron and permanent charge must be included in the equations defining the electric field, e.g., eq. (1) & eq. (6).

For physicists today, the fundamental electrical variable is the **E** field that describes the electric force on an infinitesimal test charge. **D** and **P** fields are auxiliary derived fields that many textbooks think unnecessary, at best [5-10].

## 2. Theory

Maxwell's first equation for the composite variable **D** relates the 'free charge' $\boldsymbol{\rho}_f(x, y, z|t)$, units cou/m³, to the sum of the electric field **E** and polarization **P**. It is usually written as

$$\mathbf{div}\,\mathbf{D}(x, y, z|t) = \boldsymbol{\rho}_f(x, y, z|t) \tag{1}$$

$$\mathbf{D}(x, y, z|t) \triangleq \varepsilon_0\,\mathbf{E}(x, y, z|t) + \mathbf{P}(x, y, z|t; \mathbf{E}) \tag{2}$$

The physical variable **E** that describes the electric field is not visible in the classical formulation eq. (1). Maxwell embedded polarization in the very definition of the dependent variable $\mathbf{D} \triangleq \varepsilon_0\,\mathbf{E} + \mathbf{P}$. $\varepsilon_0$ is the electrical constant, sometimes called the 'permittivity of free space'. Polarization is described by a vector field $\mathbf{P}(x, y, z|t; \mathbf{E})$ with units of dipole moment per volume, cou-m/m³, that can be misleadingly simplified to cou-m$^{-2}$. The polarization of course depends on the electric field $\mathbf{P}(x, y, z|t; \mathbf{E})$. That is why it is defined. The charge $\boldsymbol{\rho}_f$ cannot depend on **D** or **E** in traditional formulations and so $\boldsymbol{\rho}_f$ is a permanent charge.

When Maxwell's first equation is written in a style appropriate since the discovery of the electron **E** is the dependent variable, in my view. The source terms are $\boldsymbol{\rho}_f$ and the divergence of **P**.

$$\varepsilon_0 \mathbf{div}\,\mathbf{E}(x, y, z|t) = \boldsymbol{\rho}_f(x, y, z|t) - \mathbf{div}\,\mathbf{P}(x, y, z|t; \mathbf{E}) \tag{3}$$

---

[1] Thomson's monograph [3] "intended as a sequel to Professor Clerk-Maxwell's Treatise on electricity and magnetism" does not mention charge, as far as I can tell. Faraday's chemical law of electrolysis was not known and so the chemist's 'electron' postulated by Richard Laming and defined by George Stoney [4] was not accepted in Cambridge as permanent charge, independent of the electric field. It is surprising that the physical unit 'the Faraday' describes a quantity of charged particles unknown to Michael Faraday. Indeed, he did not anticipate the existence or importance of permanent charge in any form.

**P** does not have the units of charge and should not be called the 'polarization charge'. **P** does not enter the equation by itself. Only the divergence of **P** appears on the right hand side of eq (3).

$\mathbf{D}(x,y,z|t)$ and the polarization $\mathbf{P}(x,y,z|t;\mathbf{E})$ are customarily over-approximated in classical presentations of Maxwell's equations: the polarization is assumed to be proportional to the electric field, independent of time.

$$\mathbf{P}(x,y,z|t;\mathbf{E}) \triangleq (\varepsilon_r - 1)\varepsilon_0 \, \mathbf{E}(x,y,z|t) \tag{4}$$

$$\mathbf{D}(x,y,z|t) \triangleq \varepsilon_r \varepsilon_0 \mathbf{E}(x,y,z|t) \tag{5}$$

The proportionality constant $(\varepsilon_r - 1)\varepsilon_0$ involves the dielectric constant $\varepsilon_r$ which must be a single real positive number if the classical form of the Maxwell equations is taken as an exact mathematical statement of a system of partial differential equations. If $\varepsilon_r$ is generalized to depend on time, or frequency, or the electric field, the form of the Maxwell equations change. If $\varepsilon_r$ is generalized, traditional equations cannot be taken literally as a mathematical statement of a boundary value problem. They must be changed to accommodate the generalization. Many of the most interesting applications of electrodynamics arise from the nonlinear dependence of polarization and an effective $\varepsilon_r$ on field strength. In those cases, a complete model combining material and electrodynamics is needed in my opinion. Examples are presented later in this paper, near the end of Discussion.

Polarization depends on time or frequency in complex ways in all matter as documented in innumerable experiments. The frequency dependence is usually described by a generalized effective dielectric coefficient $\tilde{\varepsilon}_r$ that is not a single real number [11-16].

$\varepsilon_r$ should be taken as a constant only when experimental estimates, or theoretical models are not available, in my view, given the nearly universal complex dependence of polarization on time or frequency.

It is difficult to imagine a physical system in which the electric field produces a change in charge distribution independent of time (see examples shown towards the end of Discussion). The time range in which Maxwell's equations are used in the technology of our computers, smartphones, and video displays starts around $10^{-10}$ sec. The time range in which Maxwell's equations are used in biology start around $10^{-15}$ sec in simulations of the atoms that control protein function. The time range of the x-rays that determine protein structure is $\sim 10^{-19}$ sec. The time range used to design and operate the synchrotrons that generate x-rays is very much faster than that, something like $10^{-23}$ sec. The Maxwell equations describe experiments to many significant figures over this entire range.

It is evident that a dielectric constant $\varepsilon_r$ independent of time is an inadequate over-approximation in many cases of practical interest today, in biology, engineering, chemistry, and physics.

Despite these difficulties, **Maxwell's first equation** for **E**

$$\varepsilon_r \varepsilon_0 \mathbf{div}\, \mathbf{E}(x,y,z|t) = \boldsymbol{\rho}_f(x,y,z|t) \tag{6}$$

is often written using the dielectric constant $\varepsilon_r$ to describe polarization, without mention of the over-approximation involved. Students are then often unaware of the over-approximation, particularly if they have a stronger background in biology or mathematics than the physical sciences.

Ambiguity and its problems can be avoided if Maxwell's First Equation is rewritten without a polarization field $\mathbf{P}(x, y, z|t; \mathbf{E})$. The phenomena of polarization—the response of charges to an electric field—is then included in a variable $\boldsymbol{\rho}_Q(x, y, z|t; \mathbf{E})$

$$\boldsymbol{\rho}_Q(x, y, z|t; \mathbf{E}) \triangleq \boldsymbol{\rho}_f(x, y, z|t) - \mathbf{div}\, \mathbf{P}(x, y, z|t; \mathbf{E}) \tag{7}$$

**Maxwell's First Equation**

$$\mathbf{div}\, \varepsilon_0 \mathbf{E}(x, y, z|t) = \boldsymbol{\rho}_Q(x, y, z|t; \mathbf{E}) \tag{8}$$

Here $\boldsymbol{\rho}_Q(x, y, z|t; \mathbf{E})$ describes all charge whatsoever, no matter how small or fast or transient, including what is usually called dielectric charge and permanent charge, as well as charges driven by other fields, like convection, diffusion or temperature. The charge $\boldsymbol{\rho}_Q$ can be parsed into components in many ways, exhaustively described in [12, 13, 17-23]). Updated formulations of the Maxwell equations [13, 22] are needed, in my opinion, to avoid the problems produced by ambiguous $\mathbf{P}$ and over-simplified $\varepsilon_r$.

We adopt this version of Maxwell's first equation here.

## 3. Results

The traditional formulation shown in equations (1) and (6) are ambiguous in an important way. They do not mention the shape or boundaries of the regions in question. In fact, if $\mathbf{P}$ varies from region to region, but is constant within each region, charge is absent within each region: when $\mathbf{P}$ is constant, $\mathbf{div}\, \mathbf{P} = 0$. Charge accumulates only at the boundaries of the regions in which $\mathbf{P}$ is constant.

In many systems of dielectrics, including most of those described in classical textbooks, $\mathbf{P}$ is constant in each region and only the boundary values of $\mathbf{P}$ and (more generally charge) have effects on the Maxwell equations (1) and (6). The $\mathbf{div}\, \mathbf{P}$ field in the Maxwell equation (7), and implied in eq. (1) & (6), is zero; only the boundary values of $\mathbf{P}$ are important. The boundary values are not themselves visible (or implied) in the Maxwell equations (7), eq. (1) or (6). Nor are the boundary conditions that help determine the boundary charges evident in those equations. (Different physics and thus boundary conditions are compatible with the Maxwell equations (1), (6) and (7), and these can produce different boundary charges from the dielectric boundary conditions typically used in textbooks.) It is clear that the Maxwell equations (1), (6) and (7), in themselves do not uniquely specify the boundary charge. This ambiguity is important because the different boundary charges and physics of the boundaries of polarizable materials have an important role in the history of electrodynamics and in many applications.

Boundary conditions are particularly important when describing macromolecules in ionic solutions, like proteins in biological cells and systems. The surface of these proteins is often studded with side chains of amino acids that have permanent charge, like the carboxylate group of aspartate or glutamate side chains, and the ammonium group or guanidinium group at the end of lysine or arginine side chains. These cannot be represented by a Dirichlet boundary condition on electrical potential [24] because the proteins do not have access to sources of charge and energy to maintain the potential at a fixed value as compositions or concentrations of ions change, or conditions are changed by experiments (e.g., by site directed mutagenesis) or by biological systems (e.g., phosphorylation). Rather they can be described as a Neumann condition (to a first approximation) because the permanent charge of the surface of the Neumann protein can maintain that condition without additional sources of energy or charge. A Robin boundary condition provides a better approximation, particularly if its coefficients can be nonlinear and time dependent functions, as they are in the Hodgkin Huxley model of cylindrical neurons [25].

Dielectric boundary charges have a particular role in biological systems involving membranes. The membrane capacitance so important in determining the electrical properties of cells, particularly cells with action potentials like nerve and muscle, is a boundary phenomenon. Boundary charges are of great importance in channel proteins that allow (nearly catalyze [26]) ion flow through membranes, see Appendix on Proteins and [27].

Most of the properties of dielectric rods studied by Faraday—and predecessors going back to Benjamin Franklin, if not earlier—arise from the dielectric boundary charges. Textbooks typically spend much effort teaching why polarization charge appears on dielectric boundaries in systems with constant $\mathbf{P}$ where $\mathbf{div\,P} = 0$ (e.g., Ch. 6 of [7]). Students wonder why regions of dielectrics without polarization charge have polarization charge on boundaries.

A general principle is at work here: a field equation in itself—like eq. (1) and (6) that are partial differential equations without boundary conditions—is altogether insufficient to specify an electric field. A model is needed that has boundary conditions. The model needs to include an explicit structure. It needs to describe the spatial variation of $\mathbf{P}$. Without specifying boundary conditions (defined explicitly in specific structures), using $\mathbf{P}$ in eq. (7), and implied in eq. (1) & (6), is ambiguous and confusing. Indeed, using $\mathbf{P}$ without boundary conditions is so incomplete that it might be called incorrect.

The general nature of the ambiguity in $\mathbf{P}$ becomes clear once one realizes that:

$$\text{Adding } \mathbf{curl\,} \widetilde{\mathbb{C}}(x,y,z|t) \text{ to } \mathbf{P}(x,y,z|t) \tag{9}$$

in Maxwell's first equation, eq. (7) changes nothing because (see [28, 29])

$$\mathbf{div\,curl\,} \widetilde{\mathbb{C}}(x,y,z|t) \equiv 0 \text{ ;} \tag{10}$$

The ambiguity in $\mathbf{P}$ means that any model $\mathbf{P}_{model}(x,y,z|t;\mathbf{E})$ of polarization can have $\mathbf{curl\,} \widetilde{\mathbb{C}}(x,y,z|t;\mathbf{E})$ added to it, without making any change in the $\mathbf{div\,P}(x,y,z|t;\mathbf{E})$ in Maxwell's first equation (7), and implied in eq. (1) & (6). In other words, the polarization $\mathbf{div\,P}(x,y,z|t;\mathbf{E})$ in Maxwell's first equation (7), and implied in eq. (1) and (6), does not provide a unique structural model of polarization $\mathbf{P}_{model}(x,y,z|t;\mathbf{E})$. In particular a model drawn from an atomic detail structure can be modified by adding a polarization $\widetilde{\mathbb{P}}(x,y,z|t;\mathbf{E}) \triangleq \mathbf{curl\,} \widetilde{\mathbb{C}}(x,y,z|t;\mathbf{E})$ to its representation (i.e., 'drawing') of polarization without changing electrical properties at all: $\mathbf{div\,P}_{model} \equiv \mathbf{div\,}(\mathbf{P}_{model} + \widetilde{\mathbb{P}})$.

Models of the polarization $\mathbf{P}_{model}^1$ and $\mathbf{P}_{model}^2$ of the same structure written by different authors may be strikingly different but they can give the same electrical results even though the models can appear to be very different. The $\mathbf{curl\,} \widetilde{\mathbb{C}}(x,y,z|t;\mathbf{E})$ field can be quite complex and hard to recognize in a model, particularly for structural biologists who may not be comfortable with vector calculus and its **curl** and **div** operators. The two models $\mathbf{P}_{model}^1$ and $\mathbf{P}_{model}^2$ produce the same charge distribution **div** $\mathbf{P}_{model}^1$ and div $\mathbf{P}_{model}^2$ in Maxwell's first equation eq. (7) and so they cannot be distinguished by electrical measurements.

As we have seen, the **P** field is arbitrary, as certainly has been known previously. Purcell and Morin [5], see p. 500 – 507, describe structural models and ways to construct different fields $\mathbf{P}(x, y, z|t; \mathbf{E})$ from the same structure. **P** fields are not unique.

Purcell and Morin are not guilty of overstatement—indeed they may be guilty of understatement—when they say "The concept of polarization density **P** is more or less arbitrary" (slight paraphrase of [5], p. 507) and the **D** field is "is an artifice that is not, on the whole, very helpful" [5], p. 500.

The classical approach criticized by Purcell and Morin [5] does not allow unique specification of a polarization field $\mathbf{P}(x, y, z|t; \mathbf{E})$ from electrical measurements.

It seems clear that most formulations of electrodynamics of dielectrics in classical textbooks are "more or less arbitrary" and depend on an "artifice". An arbitrary artificial formulation is prone to artifact and likely to produce misunderstanding and unproductive argument: "what is the true description of a dielectric object (e.g., protein)?" is a question likely to arise and be unanswerable if the polarization field **P** is itself not unique.

The $\mathbf{P}(x, y, z|t; \mathbf{E})$ of classical theory is not a firm foundation on which to build an understanding of the structural basis of the phenomena of polarization, or the electrodynamics of matter, with problems particularly apparent in the understanding of the polarization arising from the structure of proteins (see Appendix).

Most applications of electrodynamics involve flow. Both biology and electrochemistry (batteries) scarcely exist without flow: what physical chemists call equilibrium (no flows of any kind) is hardly worth studying in biological or electrochemical systems. Unlike thermodynamics, electrodynamics nearly always involves flow.

Thus, we study the flux of charges $\boldsymbol{\rho}_Q$ as well as their density. Maxwell's second equation describes the flow of charges, electrical current, and the magnetic field. It is understandable that Maxwell—and his Cambridge contemporaries and followers—had difficulty understanding current flow when their models did not include permanent charge, electrons or their motions.

Maxwell's extension of Ampere's law describes the special properties of current flow $\mathbf{J}_{total}$ (eq. (12)) that make it so different from the flux of matter. Maxwell's field equations include the ethereal current $\varepsilon_0 \, \partial \mathbf{E}/\partial t$ that makes the equations resemble those of a perfectly incompressible fluid: the ethereal current always exists, whether matter is present or not, unlike the dielectric current $(\varepsilon_r - 1)\varepsilon_0 \, \partial \mathbf{E}/\partial t$ that exists only when matter is present.

Maxwell's field equations describe the incompressible flow $\mathbf{J}_{total}$ over the dynamic range of something like $10^{16}$ that is safely accessible within laboratories. The dynamic range of the Maxwell equations is much larger if one includes the interior of stars, and the core of galaxies in which light is known to follow the same equations of electrodynamics as in our laboratories.

Maxwell's field equations are different from material field equations (like the Navier Stokes equations) because they are meaningful and valid universally [30], both in a vacuum devoid of mass and matter and within and between the atoms of matter [13].

The ethereal current $\varepsilon_0 \partial \mathbf{E}/\partial t$ responsible for the special properties of Maxwell's equations arises from the Lorentz (un)transformation of charge. Charge does not vary with velocity, unlike mass, length, and time, all of which change dramatically as velocities approach the speed of light, strange as that seems. This topic is explained in any textbook of electrodynamics that includes special relativity. Feynman's discussion of 'The Relativity of Electric and Magnetic Fields' was an unforgettable revelation to me as a student [6], Section 13-6: an observer moving at the same speed as a stream of electrons sees zero current, but the forces measured by that observer are the same as the forces measured by an observed who is not moving at all. The moving observer describes the force as an electric field $\mathbf{E}(x, y, z|t)$. The unmoving observer describes the force as a magnetic field $\mathbf{B}(x, y, z|t)$. The observable forces are the same, whatever they are called, according to the principal and theory of relativity.[2]

Ethereal current is apparent in daylight from the sun, that fuels life on earth, and in night light from stars that fuels our dreams as it decorates the sky. The ethereal current is the term in the Maxwell equations that produces propagating waves in a perfect vacuum like space.

Ethereal current reveals itself in magnetic forces which have no counterpart in material fields. Magnetism $\mathbf{B}$ is given by Maxwell's form of Ampere's Law,

**Maxwell's Second Equation**

$$\frac{1}{\mu_0} \text{curl } \mathbf{B} = \mathbf{J}_Q + \varepsilon_0 \frac{\partial \mathbf{E}}{\partial t} \tag{11}$$

$$\mathbf{J}_{total} \triangleq \mathbf{J}_Q + \varepsilon_0 \frac{\partial \mathbf{E}}{\partial t} \tag{12}$$

$$\frac{1}{\mu_0} \text{curl } \mathbf{B} = \mathbf{J}_{total} \tag{13}$$

If we are interested in flux and current, we must turn to Maxwell's second equation and deal explicitly with magnetism, even if magnetic fields themselves do not carry significant energy (as in almost all biological applications). Only by dealing with Maxwell's second equation can we derive conservation of total current and compare it with the conservation of charge. Indeed, the derivation of the continuity equation used here depends on equations involving the magnetic field.

Note that $\mathbf{J}_Q$ includes the movement of all charge $\boldsymbol{\rho}_Q$ with mass, no matter how small, rapid or transient. It includes the movements of charge classically approximated as the properties of ideal dielectrics. It includes all movements of charge described by $\boldsymbol{\rho}_Q(x, y, z|t; \mathbf{E})$; $\boldsymbol{\rho}_f$ is one of the components of $\boldsymbol{\rho}_Q$. Indeed, $\mathbf{J}_Q$ can be written in terms of $\mathbf{v}_Q$ the velocity of mass with charge. In simple cases, such as a plasma of ions each with charge $\mathbf{Q}_Q$

$$\mathbf{J}_Q = \mathbf{v}_Q \mathbf{Q}_Q \mathbf{N}_Q \tag{14}$$

where $\mathbf{Q}_Q$ is the charge per particle and $\mathbf{N}_Q$ is the number density of particles. In a mixture, sets of fluxes $\mathbf{J}_Q^i$, velocities $\mathbf{v}_Q^i$, charges $\mathbf{Q}_Q^i$, number densities $\mathbf{N}_Q^i$, and charge densities $\boldsymbol{\rho}_Q^i$ are needed to keep track of each elemental species $i$ of particles. Plasmas are always mixtures because they must contain both positive and negative particles to keep electrical forces within safe bounds, as determined by (approximate) global electroneutrality.

---

[2] The principal and theory of relativity are confirmed to many significant figures every day in the GPS (global positioning systems) software of the map apps on our smartphones, and in the advanced photon sources (synchrotrons) that produce x-rays to determine the structure of proteins.

In cases other than plasmas, the relationship of $\mathbf{J}_Q, \mathbf{J}_{total}$ and $\mathbf{Q}_Q$ to material properties is complex. Those relationships must be specified separately in other models. The relationship often involves convection and diffusion fields and extends over a range of scales from atomic to macroscopic, in both space and time. For example, the Maxwell equations do not describe charge and current driven by other fields, like convection, diffusion, or temperature. They do not describe constraints imposed by boundary conditions and mechanical structures. If the other fields, structures, or boundary conditions involve matter with charge, they will respond to changes in the electric field. The other fields and constraints thus contribute to the phenomena of polarization and must be included in a description of it, as we shall discuss further below in the examples shown towards the end of Discussion. The theory of complex fluids has dealt with many such cases, often with the label 'micro macro', spanning scales, connecting micro (even atomic) structures with macro phenomena.

The charge density $\boldsymbol{\rho}_Q$ and current $\mathbf{J}_{total}$ can be parsed into components in many ways, some helpful in one historical context, some in another. Ref. [12, 13, 17-23] define and explore those representations in tedious detail. Simplifying those representations led to the treatment in this paper.

Maxwell's Ampere's law eq. (11) implies two equations of great importance and generality. First, it implies a continuity equation that describes the conservation of charge with mass. The continuity equation is the relation between the flux of charge with mass and density of charge with mass.

**Derivation**: Take divergence of both sides of eq. (11), use $\mathbf{div\ curl} = \mathbf{0}$, and get [28, 29]

$$\mathbf{div\ J}_Q = \mathbf{div}\left(-\varepsilon_0 \frac{\partial \mathbf{E}}{\partial t}\right) = -\varepsilon_0 \frac{\partial}{\partial t}\ \mathbf{div\ E} \tag{15}$$

when we interchange time and spatial differentiation.

But we have a relation between $\mathbf{div\ E}$ and charge $\boldsymbol{\rho}_Q$ from Maxwell's first equation, eq. (8), giving the Maxwell Continuity Equation:

### Maxwell Continuity Equation

$$\mathbf{div\ J}_Q = -\varepsilon_0\varepsilon_0 \frac{\partial \boldsymbol{\rho}_Q}{\partial t} \tag{16}$$

$$\mathbf{div}\left(\mathbf{v}_Q \mathbf{Q}_Q \mathbf{N}_Q\right) = -\varepsilon_0\ \frac{\partial \boldsymbol{\rho}_Q}{\partial t}, \tag{17}$$

for a biophysical or astrophysical plasma of ions

Note that sets of fluxes $\mathbf{J}_Q^i$ and sets of charge densities $\boldsymbol{\rho}_Q^i$ are needed to keep track of each elemental species $i$ of particles in a mixture, along with sets of velocities $\mathbf{v}_Q^i$, charges $\mathbf{Q}_Q^i$, and number densities $\mathbf{N}_Q^i$, as described near eq. (14).

Maxwell's Ampere's law eq. (11) implies a second equation of great importance. Indeed, it is this equation that allows the design of the one dimensional branched circuits of our digital technology using the relatively simple mathematics of Kirchhoff's current law.

**Derivation**: Taking the divergence of both sides of Maxwell's Second law eq. (11) yields

### Conservation of Total Current

$$\mathbf{div}\,\mathbf{J}_{total} = 0 \tag{18}$$

$$\mathbf{div}\,\mathbf{J}_{total} \triangleq \mathbf{div}\left(\mathbf{J}_Q + \varepsilon_0 \frac{\partial \mathbf{E}}{\partial t}\right) = 0 \tag{19}$$

or

$$\mathbf{div}\,\mathbf{J}_{total} \triangleq \mathbf{div}\left(\mathbf{v}_Q \mathbf{Q}_Q \mathbf{N}_Q \mathbf{J}_Q + \varepsilon_0 \frac{\partial \mathbf{E}}{\partial t}\right) = 0 \tag{20}$$

One dimensional systems are of great importance despite, or because of their simplicity. The design of one dimensional systems is relatively easy. It requires Kirchhoff's law and little else. The dimensionality of these circuits rules out spatial singularities. Systems are more robust when steep slopes near infinities are not present to create severe sensitivity.

Branched one dimensional systems describe most of the electronic networks and circuits of our technology. Branched one dimensional systems describe the metabolic pathways of biological cells that make life possible. Branched one dimensional systems can be described accurately by a simple generalization of Kirchhoff's law: all the $\mathbf{J}_{total}$ that flows into a node must flow out [19-22].

Unbranched one dimensional systems are also important despite their utter simplicity. Indeed, the ion channels of biological systems are unbranched one dimensional series systems. They control a wide range of biological function and cannot be considered degenerate cases. Nor can be the diodes of electronic technology that are also series systems.

Unbranched one dimensional systems have components in series, each with its own current voltage relation arising from its microphysics. In a series one dimensional system, the total current $\mathbf{J}_{total}$ is equal everywhere at any time in every location no matter what the microphysics of the flux $\mathbf{J}_Q$ of charge with mass. Maxwell's equations ensure that $\varepsilon_0\,\partial \mathbf{E}/\partial t$, and the other dependent variables, take on the values at every location and every time needed to make the total currents $\mathbf{J}_{total}$ equal everywhere. A practical example, not difficult to build in any laboratory, including resistor, capacitor, diode, capacitor, cylinder of salt water, and wire is described in detail near Fig. 2 of [20].

There is no spatial dependence of total current in a series one dimensional system. No spatial variable or derivative is needed to describe total current in such a system [23], although of course spatial variables are needed to describe other variables, including

(1) density of mass with charge $\mathbf{Q}_Q^i$.
(2) flux $\mathbf{J}_Q$ of charge with mass.
(3) electrical current $\mathbf{J}_{total}^i$ of individual elemental species.
(4) velocities, charge, and number densities $\mathbf{v}_Q, \mathbf{Q}_Q, \boldsymbol{\rho}_Q,$ and $\mathbf{N}_Q$.

It is important to realize that the flux $\mathbf{J}_Q$ of charges (with mass) can accumulate as the charge $\boldsymbol{\rho}_Q$. The flux of charge with mass $\mathbf{J}_Q$ is not conserved. In fact, $\mathbf{div}\,\mathbf{J}_Q = \mathbf{div}\,(\mathbf{v}_Q \mathbf{Q}_Q \mathbf{N}_Q)$ supplies the flow of $\mathbf{N}_Q$ plasma charges that are the current $\partial \boldsymbol{\rho}_Q/\partial t$ necessary to change $\mathbf{div}\,(\varepsilon_0\, \partial \mathbf{E}/\partial t)$, as described by the following continuity equation.

$$\mathbf{div}\,\mathbf{J}_Q = \mathbf{div}\,\varepsilon_0 \frac{\partial \mathbf{E}}{\partial t} = \frac{\partial}{\partial t}\mathbf{div}\,(\varepsilon_0 \mathbf{E}) = \frac{\partial \boldsymbol{\rho}_Q}{\partial t} = \frac{\partial\left(\mathbf{v}_Q \mathbf{Q}_Q \mathbf{N}_Q\right)}{\partial t} \tag{21}$$

Total current $\mathbf{J}_{total}$ cannot accumulate, not at all, not anywhere, not at any time but the flux of charges $\mathbf{J}_Q$ does accumulate.

Because conservation of total current applies on every time and space scale, including those of thermal motion, the properties of $\mathbf{J}_Q$ differ a great deal from the properties of $\mathbf{J}_{total}$. For example, in one dimensional channels, the material flux $\mathbf{J}_Q$ can exhibit all the complexities of a function of infinite variation, like a trajectory of a Brownian stochastic process, that reverses direction an uncountably infinite number of times in any interval, and so the Brownian trajectory is a continuous function that does not have a (well defined) time derivative anywhere. In marked contrast, the electrical current $\mathbf{J}_{total}$ has no spatial variation at all. It is spatially uniform [23]. The fluctuations of $\varepsilon_0\,\partial \mathbf{E}/\partial t$ and other variables are exactly what are needed to completely smooth the infinite fluctuations of $\mathbf{J}_Q$ into the spatially uniform $\mathbf{J}_{total}$ as strange as that seems.

## 4. Discussion

A fundamental question arises with the updated version of Maxwell's equations. How is the phenomenon of polarization included in eq. (8) & eq. (13)?

To answer this question, we first need a general paradigm to define polarization, even when dielectrics are far from ideal, when they might be time and frequency dependent, and voltage dependent as well. We need a paradigm that describes how the charge distribution varies with the electric field in as general a system as possible, including systems with charge movement driven by forces not in the Maxwell equations at all, like convection and diffusion.

This problem has been addressed in membrane biophysics. A community of scholars has studied the nonlinear currents that control the opening of voltage sensitive protein channels for nearly fifty years, inspired by [25]. They have developed protocols that may be useful in other systems, as they have been in biophysics. Schneider and Chandler and Bezanilla and Armstrong are responsible for this paradigm, more than anyone else [31-33].

I propose adopting the operational definition of 'gating current' used to define nonlinear time and voltage dependent polarization by biophysicists since 1972 [31-33]. The basic idea is to apply a set of step functions of potential and observe the currents that flow. The currents observed are transients that decline to a steady value, often to near zero after a reasonable (biologically relevant) time. The measured currents are perfectly reproducible. If a pulse is applied, the charge moved (the integral of the current) can be measured when the voltage step is applied. The integration goes on until $t_1$ when the current $i_{leak}$ is nearly independent of time, often nearly zero. That integral is called the ON charge $\mathbf{Q_{ON}}$.

When the voltage is returned to its initial value (the value that was present before the ON pulse), another current is observed that often has quite different time course [31-33], much more so than in Fig. 1. The integral of that current is the OFF charge $Q_{OFF}$.

This gating current depends on the voltage before the step. It also depends separately on the voltage after the step, although Fig. 1 does not illustrate the dependence documented in the literature [31-33]. The voltage and time dependence arises from the molecular motions underlying the gating current. The voltage and time dependence defines the mean molecular motions and is called 'the gating current' in the biophysics literature [34-36].

If the ON charge is found experimentally to equal the OFF charge, for a variety of pulse

**Figure 1**

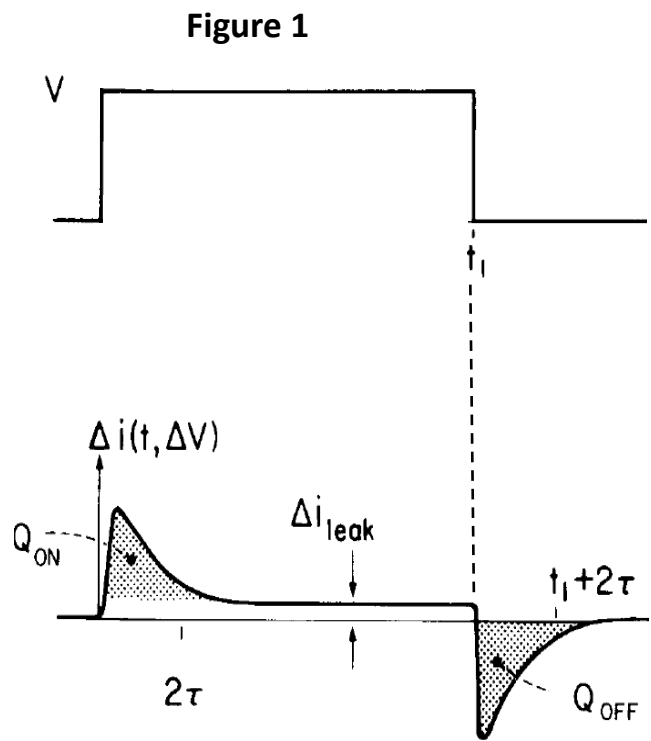

Fig. 1 shows the response to a step function change in potential and the charges measured that are proposed as an operational definition of polarization.

sizes and range of experimental conditions, the current is said to arise in a nonlinear (i.e., voltage dependent) polarization capacitance and is interpreted as the movement of charged groups in the electric field. The charged groups move to one location during the ON pulse, and return to their original location following the OFF transition. The charge is called 'gating charge' and the current that carries the charge is called 'gating current'.

The macroscale current observed in the set-up is equal to the sum of the micro (actually atomic scale) currents carried by the charged groups inside a channel protein, even though the recording electrodes are remote from the protein. Indeed, there might be $10^{18}$ charged atoms (ions) between the electrodes and the protein.

The currents in the electrodes and the channel protein are equal because the setup is designed to be an unbranched one dimensional circuit with everything in series. In a one dimensional series setup the total current is equal everywhere in the series system at any one time, even though the total current varies significantly with time. The equality of current can be checked by measuring current in different locations in the experiment. The spatial equality of current needs also to be checked in simulations as in [34-36] because tiny inadvertent errors in numerical procedures or coding can produce substantial deviations from spatial equality and thus misleading artifacts. Imposing periodic boundary conditions on nonperiodic systems is another possible source of such artifacts.

If the currents reach a steady value independent of time, but not equal to zero, as in Fig. 1, the steady current $i_{leak}$ is considered to flow in a resistive path that is time independent, but perhaps voltage dependent, in parallel with the path or device in which the gating charges $\mathbf{Q_{ON}}$ and $\mathbf{Q_{OFF}}$ flow. If the current does not reach a steady value, or if the areas are not equal, the currents are not considered 'capacitive' and are interpreted as those through a time and voltage dependent 'resistor'. It is important to check the currents through the resistive path by independent methods to see if they are time independent. In biophysics, checking can be done by blocking the resistive path with drugs, or with mutations of the channel protein. If the resistive currents depend on time, the definition of $\mathbf{Q_{ON}}$ and $\mathbf{Q_{OFF}}$ in Fig. 1 needs to be changed. Indeed, experiments of another type must be designed that allow separation of polarization from time dependent conduction currents that might mimic polarization currents.

Clearly, this approach will only work if step functions can reveal all the properties of the underlying mechanism. If the underlying mechanisms depend on the time rate of change of voltage, step functions are clearly insufficient because $\partial V/\partial t$, is zero or infinity but nothing else in a step function. In the classical language of membrane biophysics, the ionic conductances $g_{Na}$ and $g_K$ must not depend on the rate of change of voltage.

Much work has been done showing that step functions are enough to understand the voltage dependent mechanisms in the classical action potential of the squid axon [37-39], starting with [40], Fig. 10 and eq. 11. Hodgkin kindly explained the significance of this issue to colleagues, including the author (around 1970). He explained the possible incompleteness of step function measurements: if sodium conductance had a significant dependence on $\partial V/\partial t$, the action potential computed from voltage clamp data would differ from experimental measurements. He mentioned that this possibility was an important motivation for Huxley's heroic hand integration [25] of the Hodgkin Huxley differential equations. Huxley confirmed this in a separate personal communication, Huxley to Eisenberg. Those computations and many papers since [37-39] have shown that voltage clamp data (in response to steps) is enough to predict the shape and propagation of the action potential in nerve and skeletal muscle. It should be clearly understood that such a result is not available for biological systems in which the influx of Ca++ drives the action potential and its propagation [41].

The conductance of the voltage activated calcium channel has complex dependence on the current through the channel because the concentration of Ca++ in the cytoplasm is so small (~$10^{-8}$M at rest) that the current almost always changes the local concentration in the channel near the cytoplasmic side of the channel. Current through the channel changes the local concentration in and near the channel. Those concentration changes in turn alter the gating and selectivity characteristics of the channel protein, as calcium ions do in many physical and biological systems, particularly interfaces.

It seems unlikely that the resulting properties of voltage dependent calcium channels can be comfortably described by the same formalism [25] used for voltage controlled sodium and potassium channels of nerve and skeletal muscle. That formalism uses variables that depend on membrane potential and not membrane current because Cole [42] and Hodgkin [43-45] guessed that neuronal action potentials were essentially voltage dependent, not current dependent. They did not know of action potentials driven by calcium channels, nor of the extraordinarily small concentration of calcium ions inside cells. There may of course be other reasons the formalism [25] is inadequate. In summary, experiments are needed to show that responses to steps of voltage allow computation of a calcium driven action potential.

The polarization protocol described here can be applied to simulations of polarization as well as experimental measurements of polarization. Indeed, the operational definition of polarization has been applied even when theories [34] or simulations are enormously complicated by atomic detail that includes the individual motions of thousands of atoms [35, 36].

Another question of general interest is how does the polarization defined this way correspond to the polarization $\mathbf{P}$ in the classical formulation of the Maxwell equations (7), and implied in eq. (1) & (6)? Does the estimated polarization equal $\mathbf{P}$?

The answer is not pleasing. Polarization cannot be defined in general. The variety of possible responses of matter to a step of potential prevents a general answer. Indeed, a main point of this paper is that polarization must be defined by a protocol in a specific setting that specifies how the local electric field changes the distribution of charge.

Polarization cannot be defined in general because there are too many possible motions of mass with charge in response to a change in the electric field. Every possible motion of mass (with charge), including rotations and translations and changes of shape, would produce a polarization. Polarization currents can be as complicated as the motions of matter.

Insight can be developed into various kinds of polarization by constructing 'toy' models of simple systems. Those models must specify the mechanical variables $\mathbf{v}_Q, \mathbf{Q}_Q, \mathbf{\rho}_Q$, and $\mathbf{N}_Q$ (or their equivalent) and solve the field equations of mechanics, perhaps including diffusion, along with the Maxwell equations. The models are then studied using the operational definition of polarization, described previously (Fig. 1). One can hope some of the models resemble some of the more elaborate models of polarization already in the literature [16, 46-51].

Toy models might include
(1) simple electro-mechanical models, like a charged mass on a spring with damping.
(2) ideal gases of permanently charged particles, i.e., biological and physical plasmas.
(3) ideal gases of dipoles (point and macroscopic), quadrupoles, and mixtures of dipoles and quadrupoles, that rotate and translate while some are attached by bonds that vibrate (see (1)). These mixtures should provide decent representations of liquid water in ionic solutions, if they include a background dielectric, even if the dielectric is over-approximated with a single dielectric constant $\varepsilon_r(\text{H}_2\text{O}) \cong 80$.

(4) molecular models of ionic solutions that include water as a molecule. It is best to use models that are successful in predicting the activity of solutions of diverse composition and content and include water and ions as molecules of unequal nonzero size [52].

(5) classical models of impedance, dielectric, and molecular spectroscopy [16, 46-51].

(6) well studied systems of complex fluids, spanning scales, connecting micro (even atomic) structures with macroscopic functions, often called 'micro-macro models' in the literature.

These examples, taken together, will help form a handbook of practical examples closely related to the classical approximations of dielectrics.

These problems have time dependent solutions except in degenerate, uninteresting cases. Time dependence poses particular problems for the classical formulations of Maxwell equations. As stated in [22] on p. 13

"It is necessary also to reiterate that $\varepsilon_r$ is a single, real positive constant in Maxwell's equations as he wrote them and as they have been stated in many textbooks since then, following [53-55]. If one wishes to generalize $\varepsilon_r$ so that it more realistically describes the properties of matter, one must actually change the differential equation (6) and the set of Maxwell's equations as a whole. If, to cite a common (but not universal) example, $\varepsilon_r$ is to be generalized to a time dependent function, (because polarization current in this case is a time dependent solution of a linear, often constant coefficient, differential equation that depends only on the local electric field), the mathematical structure of Maxwell's equations changes.

[Perhaps it is tempting to take a short cut by simply converting $\varepsilon_r$ into a function of time $\varepsilon_r(t)$ in Maxwell's equations, as classically written.] Solving the equations with a constant $\varepsilon_r$ and then letting $\varepsilon_r$ become a function of time creates a mathematical chimera that is not correct. The chimera is not a solution of the equations. [The full functional form, or differential equation for $\varepsilon_r(t)$ must be written and solved together with the Maxwell equations. This is a formidable task in any case, but becomes a challenge more than a task if convection or electrodiffusion modify polarization, as well as the electric field.]

If one confines oneself to sinusoidal systems (as in classical impedance or dielectric spectroscopy [11, 46, 56, 57]), one should explicitly introduce the sinusoids into the equations and not just assume that the simplified treatment of sinusoids in elementary circuit theory [58-62] is correct: it is not at all clear that Maxwell's equations always have steady state solutions in the sinusoidal case when combined with other field equations (like Navier Stokes [63-80] or PNP = drift diffusion [68, 81-96]); [joined] with constitutive equations; and boundary conditions. They certainly do not always have solutions that are linear functions of just the electric field [97-100]."

It seems clear that the classical Maxwell equations with the over-approximated dielectric coefficient $\varepsilon_r$ cannot emerge in the time dependent case. Of course, the classical Maxwell equations cannot emerge when polarization has a nonlinear dependence on the electric field, or depends on the global (not local) electric field, or depends on convection or electrodiffusion.

Indeed, in my opinion, when confronted with the models of polarization listed on the previous page, the classical Maxwell equations will be useful only when knowledge of the actual properties of polarization is not available. All the models listed involve time dependence in the polarization fields that are not included in the classical Maxwell equations as usually written.

## 5. Conclusions

A generalization of Maxwell's **P** useful in a range of systems may emerge. The generalization would describe how the local electric field changes the distribution of charge, as one imagines that Maxwell hoped **P** and **D** would be.

Until then, one is left with

(1) bewilderingly complete measurements, over an enormous range of frequencies (e.g., [14, 16, 46-51]) of the dielectric properties and conductance of ionic solutions of varying composition and content. These measurements embarrass the theoretician with their diversity and complexity. They have not yet been captured in any formulas or programs less complicated than a look-up table of all the results.

(2) computations of the motion of all charges on the atomic scale [35, 36], described by the field equations of mechanics and electrodynamics [34].

(3) reduced models. It is unlikely that the reduced models can be derived solely by mathematics. It is more likely that they must be 'guessed and checked' one by one, as most models are checked in science.

Sadly, the actual properties of polarization are often unknown. Then, one is left with the over-approximated eq. (6) or nothing at all. It is almost never wise to assume polarization effects are negligible. Eq. (6) is certainly better than nothing. Eq. (6) can be particularly helpful if it is used gingerly: toy models can successfully represent an idealized view of a part of the real world of technological or biological importance, for example, electronic circuits or several properties of ion channels.

In some cases, the toy models can be enormously helpful. They allow the design of circuits in our analog and digital electronic technology [101-104]. They allow the understanding of selectivity [52, 105-107] and current voltage relations of several important biological channel proteins in a wide range of solutions [52, 108-110]. In other cases—for example the description of ionic solutions with many components—the toy models can be too unrealistic to be useful. Experiments and experience can tell how useful the toy model actually is in a particular case: pure thought usually cannot.

## Acknowledgment

It is a particular pleasure to thank my friend and teacher Chun Liu for his continual encouragement and advice, and for patiently correcting mistakes in my mathematics as these ideas were developed over many years. Mistakes may remain, sad to say. All are my responsibility.

# Appendix: $P(x, y, z|t; E)$ in Proteins

Ambiguities in the meaning of the polarization field $P(x,y,z|t; E)$ can cause serious difficulties in the understanding of protein function. Understanding protein function is greatly aided by knowledge of protein structure. The protein data bank contains 168,095 structures in atomic detail today (August 27, 2020) and the number is growing rapidly as cryo-electron microscopy is used more and more.

Protein structures are usually analyzed with molecular dynamics programs that assume periodic boundary conditions and chemical equilibrium, i.e., no flows. Most proteins control large flows as part of their natural biological functions. Equilibrium hardly ever occurs in living biological systems. It seems obvious that equilibrium systems cannot provide general insight into flows, any more than a nonfunctional amplifier without a power supply can show how a functional amplifier works. Proteins are not periodic in their natural setting. It seems obvious that periodic systems with flow cannot conserve total current $\mathbf{J}_{total}$ in general—or perhaps even in particular—as required by the Maxwell equations, see eq. (19). In other words, it is likely that molecular dynamics analyses of periodic structures do not satisfy the Maxwell equations, although almost all known physics does satisfy those equations.

It is also unlikely that standard programs of molecular dynamics compute electrodynamics of nonperiodic systems correctly, despite their use of Ewald sums, with various conventions, and force fields (tailored to fit macroscopic, not quantum mechanical) data. Compare the exhaustive methods used to validate results in computational electronics [88] with those in the computation of electric fields in proteins.

The electrostatic and electrodynamic properties of proteins are of great importance. Many of the atoms in a protein are assigned permanent charge greater than $0.2\mathbf{e}$ in the force fields used in molecular dynamics, where $\mathbf{e}$ is the elementary charge, and these charges tend to cluster in locations most important for biological function, just as they cluster at high density near the electrodes of batteries and other electrochemical systems. Enormous densities of charge ($> 10M$, sometimes much larger) are found in and near channels of proteins [24, 52, 111, 112] and in the 'catalytic active sites' [113] of enzymes. Such densities are also found near nucleic acids, DNA and (all types of) RNA and binding sites of proteins in general.

A feel for the size of electrostatic energies can be found from Coulomb's law between isolated charges (in an infinite domain without boundary conditions).

$$E_{cou} = \frac{560}{\varepsilon_r} \frac{q_i q_j}{r_{ij}} \tag{22}$$

Here $E_{cou}$ is in units of the thermal energy $RT$, with gas constant $R$ and absolute temperature $T$; charge $q_i$ or $q_j$ are in units of the elementary charge $\mathbf{e}$; and $r_{ij}$ is in units of Angstroms $= 10^{-10}$ meters.

For water, with $\varepsilon_r \cong 80$, this becomes

$$E_{cou}(\text{water}) = 7 \frac{q_i q_j}{r_{ij}}; \text{ units } RT \tag{23}$$

For the pore of a channel, one can guess $\varepsilon_r \cong 10$ and then

$$E_{cou}(\text{channel: } \varepsilon_r \cong 10) = 56 \frac{q_i q_j}{r_{ij}}; \text{ units } RT \tag{24}$$

Electrostatic energy has to be computed very accurately indeed to predict current voltage relations observed experimentally. Gillespie [110] reported mostly in 'Supplemental Data' that errors of energy of 0.05 RT produced significant changes in current voltage relations of the ryanodine receptor protein. Similar sensitivity is expected near nucleic acids, or for any ion channel or active site of an enzyme or binding site of a protein because the underlying energetics are similar.

Accuracy of this sort is not claimed in most calculations of molecular or quantum dynamics involving ionic solutions. Typical accuracies claimed for quantum chemical calculations start around 2 RT and for molecular dynamics calculations around 0.5 RT, in my (limited) experience. However, reduced models of proteins allow calculations of the precision needed to deal with Gillespie's results, as much work demonstrates [108, 110, 114-117].

Reduced models use lower resolution representations drawn from the full detail atomic structure. Typically the accuracy of the reduced model itself cannot be calculated from first principles, but the model can be checked to be sure it conserves mass, charge and total current, on all scales. If the model fits a wide range of data, measured in solutions of different composition and contents with a single set of parameters, as does Gillespie's, and it satisfies conservation laws, the precision of the model is assured by the fit to the data itself [110].

Polarization effects are important in proteins as well as channels. The enormous effort and investment in developing polarizable force fields [118] is an eloquent testimonial to the importance of polarization in biological and chemical applications. Charges that depend on the electric field are as important as the permanent charges analyzed in the last paragraphs (see [119] and references cited there).

The ambiguous nature of the **P** field means that **P** provides a poor guide to the importance of polarization in a particular protein. Identical electrodynamics will arise from structures that appear different, but only differ by $\widetilde{\mathbb{P}}(x,y,z|t) \triangleq \mathbf{curl}\ \widetilde{\mathbb{C}}(x,y,z|t)$. Situations like this can produce confusion and unproductive argument, because not all scientists know of the inherent unavoidable ambiguity of **div P** or (7), and implied in eq. (1) & (6),: $\mathbf{div}\ \mathbf{P} \equiv \mathbf{div}\left(\mathbf{P} + \widetilde{\mathbb{P}}\right)$.

What is needed is a model of both the dynamics of mass and the electrodynamics of charge, that allows the unambiguous calculation of the response of the protein to an applied electric field. The combined dynamics of mass and electrodynamics of charge are the appropriate model of the polarization phenomena, not the classical **P** field.

Low resolution models may do surprisingly well provided they
    (1) actually allow flow
    (2) satisfy the Maxwell equations and conservation of mass, charge, and total current
    (3) avoid periodic boundary conditions for the electric field in nonperiodic systems.

Compare [34] and [35, 120].